\documentclass[12pt]{AIAA}
\usepackage{amsmath,amssymb,amsthm}
\usepackage{graphicx}
\newcommand{\R}{\mathbb{R}}

\newcommand{\Sone}{\mathbb{S}^1}
\def\SO3{\mathop{\mathbb{SO}(3)}}
\def\so3{\mathfrak{so(3)}}

\newcommand{\Irotord}{\mathbb{I}_{r}}

\newcommand{\Igimbal}{\mathbb{I}_{g}}
\newcommand{\Igimrot}{\mathbb{I}_{gr}}
\newcommand{\Is}{\mathbb{I}_{s}}
\newcommand{\Itot}{\mathbb{I}_{total}}
\newcommand{\Itilde}{\tilde{\mathbb{I}}(x)}

\def\rta{\mathop{\rightarrow}}
\def\deff{\stackrel{\triangle}{=}}
\newcommand{\pmat}[1]{\begin{pmatrix}#1\end{pmatrix}}
\newcommand{\inprod}[2]{\left\langle{#1}, {#2}\right\rangle}
\def\riem{\mathbb{G}}

\def\skew#1{\mathcal{S}{(#1)}}

\def\bei{\begin{itemize}}
\def\ei{\end{itemize}}

\newtheorem{claim}{Claim}[section]
\newtheorem{definition}{Definition}[section]

\begin{document}

\title{The Principal Fiber Bundle Structure of the Gimbal-Spacecraft System}

\author{Ravi N. Banavar\footnote{banavar@sc.iitb.ac.in} and Arjun Narayanan\footnote{arjun\_n@sc.iitb.ac.in}}
\affiliation{Systems and Control Engineering, Indian Institute of Technology Bombay, Mumbai 400076, India}


\maketitle

\section*{Nomenclature}
\noindent\begin{tabular}{@{}lcp{4.5in}@{}}
$\beta$,$\gamma (\in \Sone) $ &=&
The gimbal and wheel angles. (\(rad\))\\
$R_{\beta} (\in \SO3)$ &=&
Transformation from gimbal frame $\mathcal{G}$
to the spacecraft body frame $\mathcal{B}$.\\
$\Is$ &=&
Spacecraft inertia without the CMG gimbal and wheel inertia. (\(kg.m^2\))\\
$I_{g},\,I_{r}$ &=&
Gimbal frame inertia, wheel inertia about own centre of mass represented in gimbal frame. (\(kg.m^2\))\\
$(\Igimrot)_{\beta}$ &=&
Combined inertia of gimbal frame and wheel in the spacecraft frame.
$R_{\beta}  \Igimrot R_{\beta}^T$. (\(kg.m^2\))\\
$\tilde{I}_{(\beta)}$ &=&
Locked inertia tensor.\\
$X$ &=&
State variable 
$\triangleq(R_{s},\beta,\gamma)\triangleq(R_{s},x)$.\\
$\Omega_{s}$ &=&
Angular velocity of the spacecraft in the  body frame. (\(rad/s\))\\
$\mu$ &=&
Total spatial angular momentum of the spacecraft in inertial frame. (\(kg.m^2/s\))\\
$i_{2},i_{3}$ &=&
The vectors 
$[0\,1\,0]^{T}$ and 
$[0\,0\,1]^{T}$.\\
$\mathcal{S}(\ ),\widehat{(\ )}$ &=&
Mapping 
$\mathbb{R}^{3}\rightarrow\mathfrak{so}(3)$
such that 
$\mathcal{S}(\vec{a})\vec{b}\triangleq\hat{\vec{a}}\cdot\vec{b}\triangleq\vec{a}\times\vec{b}$.
\end{tabular} \\

\section{Introduction}
Spacecrafts are actuated by two principles - internal or external actuation. The actuators in the former 
class consist of momentum wheels (also called internal rotors) and control moment gyros (CMGs).
A refinement of CMGs are the variable speed CMGs (called VSCMGs).
Examples of external actuation systems include gas jet thrusters mounted 
on the outer body of the spacecraft.
In this article we focus on spacecraft with gimbals (or the VSCMG) 
as the mechanism of actuation.

The modelling of  a spacecraft with a VSCMG has been reported in the literature
\cite{schaub_feedback_1998,schaub_singularity_2000}. 
Studies on singularity issues of this system are found in
\cite{tsiotras_singularity_2004,schaub_singularity_2000}. Control
law synthesis and avoidance of singularities are found in
\cite{bedrossian_thesis_1987,margulies_aubrun_1978,
schaub_singularity_2000,tsiotras_singularity_2004} An early 
study of singularity in a geometric framework is \cite{kurokawa_geometric_1998}.
That study considers possibility of avoiding singular
gimbal angle configurations using global control.
The analysis proceeds by considering the inverse images of
CMG system angular momenta as union of sub\emph{``manifolds''} of the n-dimensional 
gimbal angle configuration space.
The nature of these manifolds are examined at and near singular gimbal configurations.

Spurred by the insight and creativity of J. E. Marsden
\cite{marsden2013introduction} ,the geometric mechanics community
\cite{bloch_nonholo_mech_ctrl,ostrowski1999computing} has studied very
many mechanical systems in the geometric framework. This
framework has proved beneficial 
in providing insight into these systems and their structure by preserving the mechanical
objects (momentum, energy) of these systems and also proving useful in 
control design \cite{bloch2001controlledlagrangian,bloch2000controlled}.

A geometric description of the VSCMG system based on variational principles is
studied in \cite{sanyal2013vscmg}. The configuration space of the system
is shown to be a principal fiber bundle and the expression for 
the Ehressmann connection is derived.
The paper considers a general system where the rotor mass centre is
offset from the gimbal axis.
A stabilising control law is derived as a function of the
internal momentum. Singularity analysis of this system under
typical simplifying assumptions is explored in \cite{sanyal2015vscmg}.
The system is discretised into a model which preserves the conserved
quantities using variational integrators.

\begin{figure}[h!]
    \centering
    \includegraphics[scale=0.6]{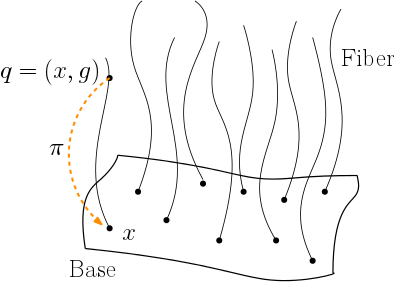}
    \caption{Fiber Bundle}
    \label{fiber_bundle}
\end{figure}

While studying the problem of interconnected mechanical systems, the geometry of
the configuration space which is a differential manifold requires attention for
elegant and insightful solutions.
This configuration space $Q$, is often written as the product of two manifolds.
One component is the base manifold $M$, which in our context describes the configuration
of the gimbals mounted
inside the spacecraft. The other 
component of the configuration variables depicting the attitude of the spacecraft is
a Lie group $G$, in this case $\SO3$ 
\cite{boothby1975introduction}. The total configuration space of the 
spacecraft $Q$ then naturally appears as a product $G \times M$.
Such systems follow the topology of a trivial principal fiber bundle,
see \cite{kobayashi1963foundations}.
Figure \ref{fiber_bundle} shows an explanatory figure of a fiber bundle. The components of
the 2-tuple
$q = (x, g)$ denote the base and the group variable respectively.  The projection map $\pi:
Q \rta M$ maps the configuration space to the base space.
With such a separation of the configuration space, locomotion is readily seen as
the means by which changes in shape affect the macro position.
We refer to \cite{kelly1995geometric,bloch_nonholo_mech_ctrl}
for a detailed explanation on the topology of locomoting systems.

In this article we present a completely geometric approach to the modelling of the 
VSCMG-spacecraft system and relate it to a conventional modelling approach.


\section{Modelling in a geometric framework}
    The configuration space of the spacecraft-gimbal system (with one CMG) is $Q = \SO3 \times \Sone \times \Sone $
    and any arbitrary configuration is expressed by the 3-tuple $(R_s, \beta, \gamma)$, where
    the first element denotes the 
    attitude $R_s$ of the spacecraft with respect to fixed / inertial frame, the second denotes the degree of 
    freedom $\beta$ of the gimbal frame, the third denotes the degree of freedom 
    $\gamma$ of the 
    rotor about its  spin axis. For the purpose of later geometrical interpretation, we club
    $x \deff (\beta, \gamma)$. There are three rigid bodies involved here, 
    each having relative motion (rotation) about the other. Therefore, three frames of reference
    are chosen (apart from the inertial frame) - the first is the 
    spacecraft, denoted by the subscript $s$, the second is the gimbal frame, denoted by the 
    subscript $g$, the third is the rotor frame, denoted by the subscript $r$.
    The moments of inertia of the homogeneous rotor and the gimbal in their respective 
    body frames are assumed to be
    \begin{align}
    \Irotord=  \pmat{J_x & 0 & 0 \\ 0 & J_x &  0  \\  0  &  0  & J_z } \;\;\;
    \Igimbal = \pmat{I_t & 0 & 0 \\ 0 & I_g &  0  \\  0  &  0  & I_s }
    \end{align}
    Since the rotor is 
    assumed to be homogeneous and symmetric, its inertia is represented in the gimbal frame  and
    the combined gimbal-rotor inertia is rewritten as     
    \begin{align}
    \Igimrot= \pmat{(J_x + I_t)& 0 & 0 \\ 0 & (J_x + I_g) &  0  \\  0  &  0  & (J_z + I_s) }
    \end{align}
    If the rotational transformation that relates the gimbal frame to the spacecraft frame is 
    given by $R_{\beta}$, where
    $\beta$ denotes the gimballing angle,
    then the gimbal-rotor inertia reflected in the spacecraft frame is
    \begin{align}
    (\Igimrot)_\beta \deff R_{\beta}  \Igimrot R_{\beta}^T
    \end{align}
    Here the subscript $\beta$ denotes the dependance on the gimbal angle $\beta$.
    \subsection{Kinetic energy and a Riemannian structure}
	     The angular velocity of the spacecraft, $\Omega_s \in \R^3$, 
	     represented as a skew-symmetric matrix $ {\hat{\Omega}_s} $, 
	     is an element of the Lie algebra
    $\so3$. The inner product on this Lie algebra $\so3$ 
    is defined  in terms of the 
    standard inner product on $\R^3$ as 
    \begin{align}
    \inprod{\hat{\Omega}_1}{\hat{\Omega}_2}_{\so3}     \deff  \frac{1}{2} \inprod{  \Omega_1}{   \Is  \Omega_2}_{\R^3}.
    \end{align}   
	     and denotes the kinetic energy of the spacecraft body.
    The kinetic energy of the gimbal-rotor unit is given by
    \begin{align}
    \frac{1}{2} \inprod{R_{\beta}^T \Omega_s + \pmat{0 \\ \dot{\beta} \\ \dot{\gamma}} }
    {\Igimrot  [ R_{\beta}^T \Omega_s + \pmat{0 \\ \dot{\beta} \\ \dot{\gamma}} ] }
    \end{align}
		and the total kinetic energy of the entire spacecraft-gimbal system is
    \begin{align}
    \frac{1}{2} \inprod{  \pmat{\Omega_s \\ 0 \\ \dot{\beta} \\ \dot{\gamma}}   }
    { \Itot   \pmat{\Omega_s \\ 0 \\ \dot{\beta} \\ \dot{\gamma}}     },
    \end{align}
    where 
    \begin{align}
    \Itot(\beta)= \pmat{ (R_{\beta}  \Igimrot R_{\beta}^T + \Is)  &  R_{\beta} \Igimrot  \\  \Igimrot  
        R_{\beta}^T &  \Igimrot   }.
    \end{align} 
    It is to be noted that the inertia matrix is dependant on the gimbal angle, $\beta$.

    The kinetic energy induces a metric on the configuration space $Q$ of the
    system, which enables us to impart a {\it Riemannian structure}
    to the system. The Riemannian metric $\riem$ defines a smoothly varying
    inner product on 
    each tangent space of $Q$.  For $q = (R_s, (\beta, \gamma)) \in Q $ and 
    $v_q = (R_s\hat{\Omega}_1, (v_\beta,v_\gamma)), 
    w_q = (R_s\hat{\Omega}_2, (w_\beta,w_\gamma)) \in T_qQ$, the Riemannian metric
    is defined as 
    \begin{align}
    \inprod{v_q}{w_q}_{\riem} & =  \riem(q) (v_q, w_q)  \nonumber \\
    & =   \riem ( R_s, (\beta, \gamma)) \left( (R_s \hat{\Omega}_1 , (v_\beta, v_\gamma)),
    (R_s \hat{\Omega}_2 , (w_\beta, w_\gamma))\right)
    \end{align}
    \begin{align}
    =  \frac{1}{2} \inprod{  \pmat{R_s \hat{\Omega}_1 \\ 0 \\ v_\beta \\ v_\gamma }   }
    { \Itot   \pmat{R_s \hat{\Omega}_2 \\ 0 \\ w_\beta \\  w_\gamma }     }                                                                             
    \end{align}
    Here $\hat{\Omega}_1$ and
    $ \hat{\Omega}_2$
    belong to the Lie algebra $\so3$. Note that we have used the left-invariant property 
    of the vector field on $\SO3$, and the fact that the Riemannian metric on $\SO3$
    is induced by the inner product on the Lie algebra $\so3$, wherein 
    \begin{align}
    \inprod{v_R}{w_R}_{\SO3} \deff \inprod{\hat{\Omega}_1}{\hat{\Omega}_2}_{\so3} ,
    \end{align}
    where $v_R = R \hat{\Omega}_1$ and $w_R = R \hat{\Omega}_2$, the left translations
    by $R$ of $\hat{\Omega}_1$ and $\hat{\Omega}_2$, respectively.
    \subsection{Group action and a principal fiber bundle}

    The action of the $\SO3$ group  on $Q$ induces more geometric structure into 
    the 	problem. 
    Given $M \in \SO3$, the action  is defined by 
    \begin{align}
    \SO3 \times Q \rta Q \;\;\;\;\;\;   (M , (R_s, \beta, \gamma))  \rta (MR_s, \beta, \gamma)
    \end{align}
    and the corresponding tangent lifted action is given by 
    \begin{align}
    T \SO3 \times TQ \rta TQ \;\;\;   (v_{R_s}, v_{\beta}, v_{\gamma})  \rta 
    (Mv_{R_s}, v_{\beta}, v_{\gamma})
    \end{align}
    This action is chosen based on the symmetry in the system;
    in this case, the fact that the kinetic energy of the spacecraft in a potential field free space
    remains unchanged under rotational transformations.
    The gimbal and the rotor configuration variables 
    are viewed in a {\it base space (or shape space)} and the rigid body 
    orientation is viewed as a group variable in a {\it fiber space}, and with a few additional
    requirements, the model is
    amenable to a principal fiber bundle description. See \cite{bullo_modelling_ctrl}
    for more details on 
    describing mechanical systems in a fiber bundle framework.
    The fiber bundle structure separates the 
    actuation and orientation variables and proves beneficial and intuitive in control design.
    
    Based on the above model description, we identify the principal fiber bundle 
    $(Q, B, \pi, G)$,
    where $Q = \SO3 \times \Sone \times \Sone$, 
    $B = \Sone \times \Sone $ and $\pi: Q \rta B$ is a bundle projection map. 
    \begin{claim}
        Under the defined group action, the kinetic energy of the total system 
        remains invariant.
    \end{claim}      
    \proof
    Straightforward.
    $\Box$.
    
    We now define
    a few geometric quantities on this fiber bundle on the lines of 
    \cite{bloch_nonholo_mech_ctrl}. 
    \bei
    \item 
    The {\it infinitesimal generator} of the Lie algebraic 
    element $\hat{\eta} \in \so3$ under the group action is the vector field 
    \begin{align}
    \hat{\eta}_{Q}(q) = \frac{d}{dt}|_{t=0} (\exp(\hat{\eta}t) R_s, (\beta, \gamma)) = 
    (\hat{\eta}R_s, (0, 0))
    \end{align}
    \item The {\it momentum map} $J : TQ \rta \so3^*$ is given by
    \begin{align}
    [J(q, v_q), \xi] = \inprod{ v_q   }{  \hat{\xi}_Q(q) }_{\riem}
    \end{align}
    and 
    since the kinetic energy is invariant under the action of the $\SO3$ group, we have
    \begin{align}
    \inprod{ v_q   }{  \hat{\xi}_Q(q) }_{\riem} = \inprod{ v_{(e,(\beta,\gamma))} }
    {(R_s^T \hat{\xi} R_s,(0,0))  }_{\riem}
    \end{align}
    which yields
    \begin{align}
    [J(q, v_q), \hat{\xi}] =  \inprod{Ad_{R_s^T}^* [( (\Igimrot)_s + \Is) \Omega_s + R_{\beta} \Igimrot \pmat{ 0  \\  \dot{\beta}
            \\  \dot{\gamma} }   ] }{\xi}
    \end{align}
    where $\inprod{\cdot}{\cdot}$ denotes the inner product, while 
    $[\cdot, \cdot]$ denotes the primal-dual action of the vector spaces $\so3$ and
    $\so3^*.$
    In the absence of external forces, the momentum map is conserved.
    Since the total spatial angular momentum of the system is constant, say $\mu$, 
    the above expression
    yields
    \begin{align}
    \mu = Ad_{R_s^T}^* [( (\Igimrot)_s + \Is) \Omega_s + R_{\beta} \Igimrot \pmat{ 0  \\  \dot{\beta}
        \\  \dot{\gamma} }   ]   
    \end{align}
    \item 
    The {\it locked inertia tensor} at each point $q \in Q$ is the mapping 
    \begin{align}  \mathbb I (q): \so3 \rta \so3^* \end{align}
    and is defined as  
    \begin{align}  [ \mathbb I(q) \eta, \xi ] = \riem (q) \left((\hat{\eta} R_s , (0,0)),
    (\hat{\xi} R_s , (0, 0))\right)
    \end{align}
    \item 
    The {\it mechanical connection} is then defined as  the $\so3$-valued one-form	
    \begin{align}
    \alpha: TQ \rta \so3 \;\;\;\; (q, v_q) \rta \alpha(q, v_q) = {\mathbb I(q)}^{-1} J(q,v_q)
    \end{align}
    \ei

           We now proceed to present a kinematic model and a dynamic model for the
    system under consideration.
    With the state-space as $X \deff (R_s, (\beta, \gamma)) = (R_s, x)$, where $x \deff (\beta, \gamma)$ 
    and defining $\Itilde =  (R_{\beta}  \Igimrot R_{\beta}^T + \Is) $, the control inputs
    (gimbal velocity and rotor spin)  at the kinematic
    level as $u \deff \dot{x} = (\dot{\beta}, \dot{\gamma}) = \pmat{u_{\beta}  \\  
        u _{\gamma} } $,
    the affine-in-the-control system model is
    \begin{align}
    \dot{X} = f(X) + g(X) u 
    \end{align}
    where the drift and control vector fields are given by 
    \begin{align}
    f(X) = \pmat{R_s  \skew{ (  \Itilde^{-1} (Ad_{R_s}^* \mu))   }    \\  0 }
    \end{align}
    \begin{align}
    g_{\beta}(X) =  \pmat{- R_s  \skew{   (\Itilde^{-1}  (Ad_{R_\beta^T}^* (\Igimrot i_2))  }   
        \\  
        \pmat{1  \\  0}   }
    \;\;\;\;\;
    g_{\gamma}(X) =  \pmat{- R_s  \skew{ (\Itilde^{-1}  (Ad_{R_\beta^T}^* (\Igimrot i_3))  }     
        \\  
        \pmat{0  \\   1}     }
    \end{align}
    Here  $\skew{\cdot} : \R^3 \rta \so3$ is given by 
    \begin{align}
    \skew{(\psi_1, \psi_2, \psi_3)} \deff 
    \pmat{ 0 &  - \psi_3  &  \psi_2  \\  \psi_3  &  0  &  -\psi_1  \\  
        - \psi_2  &  \psi_1  &  0 }
    \end{align}

\section{The dynamic model}
To arrive at the dynamic model we proceed as follows.
From the expression for the total momentum 
\begin{align}
\mu  & = Ad_{R_s^T}^* [ ((\Igimrot)_s + \Is) \Omega_s + R_{\beta} \Igimrot \pmat{ 0  \\  \dot{\beta}
    \\  \dot{\gamma} }   ] \nonumber   \\
& = R_s \pmat{ \Itilde & R_{\beta} \Igimrot}  \pmat{\Omega_s 
    \\  \pmat{ 0  \\  \dot{\beta}  \\  \dot{\gamma} }  }
\end{align}
We split the momentum in to two components - one due to the gimbal-rotor unit and the other due to
the rigid spacecraft. Further, we assume an internal torque $\tau_b$ 
(in the gimbal-rotor frame), generated by a motor, 
acts on the gimbal and rotor unit. We then have, due to the principle of action and
reaction
\begin{align}
\frac{d}{dt} (R_s \Is \Omega_s) = \underbrace{- R_s \tau_b}_{reaction}
\;\;\;\; \;\;
\frac{d}{dt} ( R_s [ R_{\beta} \Igimrot R_{\beta}^T \Omega_s + R_{\beta} \Igimrot \dot{x} ] )
= \underbrace{R_s \tau_b}_{action}
\end{align}  
The detailed computation is now shown. 
Differentiating 
$
\mu=R_{s}R_{\beta} \Igimrot (R_{\beta}^{T}\Omega_{s}+\begin{bmatrix}0\\
\dot{\beta}\\
\dot{\gamma}
\end{bmatrix})
$
with respect to time, we have
$\frac{d}{dt}\mu_{cmg}=\frac{d}{dt}\left\{ R_{s}R_{\beta} \Igimrot (R_{\beta}^{T}\Omega_{s}+\begin{bmatrix}0\\
\dot{\beta}\\
\dot{\gamma}
\end{bmatrix})\right\} =\text{torque acting on cmg external to cmg}$

$\tau_{extcmg}=\left\{ \begin{array}{c}
(R_{s}\hat{\Omega}_{s})R_{\beta} \Igimrot (R_{\beta}^{T}\Omega_{s}+\begin{bmatrix}0\\
\dot{\beta}\\
\dot{\gamma}
\end{bmatrix})\\
+R_{s}(R_{\beta}\hat{i}_{2}\dot{\beta}) \Igimrot (R_{\beta}^{T}\Omega_{s}+\begin{bmatrix}0\\
\dot{\beta}\\
\dot{\gamma}
\end{bmatrix})\\
+R_{s}R_{\beta} \Igimrot (-\hat{i}_{2}R_{\beta}^{T}\dot{\beta}\Omega_{s})\\
+R_{s}R_{\beta} \Igimrot (R_{\beta}^{T}\dot{\Omega}_{s})\\
+R_{s}R_{\beta} \Igimrot (\begin{bmatrix}0\\
\ddot{\beta}\\
\ddot{\gamma}
\end{bmatrix})
\end{array}\right\} $

\begin{align*}\tau_{extcmg}={} & R_{s}\hat{\Omega}_{s}R_{\beta} \Igimrot (R_{\beta}^{T}\Omega_{s}+\begin{bmatrix}0\\
\dot{\beta}\\
\dot{\gamma}
\end{bmatrix})\\
&+R_{s}R_{\beta}\left\{ \hat{i}_{2}\dot{\beta} \Igimrot (R_{\beta}^{T}\Omega_{s}+\begin{bmatrix}0\\
\dot{\beta}\\
\dot{\gamma}
\end{bmatrix})+ \Igimrot (-\hat{i}_{2}R_{\beta}^{T}\dot{\beta}\Omega_{s})+ \Igimrot (\begin{bmatrix}0\\
\ddot{\beta}\\
\ddot{\gamma}
\end{bmatrix})\right\}
\end{align*}

In the spacecraft body coordinates,

\begin{align*}\tau_{extcmg}^{\mathcal{B}}={}&\hat{\Omega}_{s}R_{\beta} \Igimrot (R_{\beta}^{T}\Omega_{s}+\begin{bmatrix}0\\
\dot{\beta}\\
\dot{\gamma}
\end{bmatrix})\\
&+R_{\beta}\left\{ \hat{i}_{2}\dot{\beta} \Igimrot (R_{\beta}^{T}\Omega_{s}+\begin{bmatrix}0\\
\dot{\beta}\\
\dot{\gamma}
\end{bmatrix})+ \Igimrot (-\hat{i}_{2}R_{\beta}^{T}\dot{\beta}\Omega_{s})+ \Igimrot (\begin{bmatrix}0\\
\ddot{\beta}\\
\ddot{\gamma}
\end{bmatrix})\right\}
\end{align*}
Defining two vectors
\begin{align*}
u_{1}={} &\Igimrot (R_{\beta}^{T}\Omega_{s}+\begin{bmatrix}0\\
\dot{\beta}\\
\dot{\gamma}
\end{bmatrix})\\
u_{2}={} &\left(\hat{i}_{2} \Igimrot - \Igimrot \hat{i}_{2}\right)R_{\beta}^{T}\Omega_{s}\dot{\beta}+\hat{i}_{2} \Igimrot \begin{bmatrix}0\\
\dot{\beta}\\
\dot{\gamma}
\end{bmatrix}\dot{\beta}+ \Igimrot (\begin{bmatrix}0\\
\ddot{\beta}\\
\ddot{\gamma}
\end{bmatrix})
\end{align*}
in the gimbal frame $\mathcal{G}$ such that
$\tau_{extcmg}^{\mathcal{B}}=\hat{\Omega}_{s}R_{\beta}u_{1}+R_{\beta}u_{2}$
. The second component of this vector gives the torque acting on the gimbal motor
and the third gives the torque acting on the wheel motor.
We now simplify this expression to obtain more explicit equations, which we then 
compare with a standard model existing in the literature.
\[
[ \hat{\Omega}_s \Itilde \Omega_s  +  \Itilde \dot{\Omega}_s ] 
\]
\begin{align}
= (-1)\hat{\Omega}_s R_{\beta} \Igimrot \dot{x} - \dot{\beta} R_{\beta} {\cal U} R_{\beta}^T
\Omega_s - \dot{\beta} R_{\beta} \hat{i}_2 \Igimrot \dot{x} - R_{\beta} \Igimrot 
\ddot{x}
\end{align}
where ${\cal U} \deff \hat{i}_2 \Igimrot - \Igimrot \hat{i}_2$ is a symmetric matrix.

\section{Comparison to the Schaub-Rao-Junkins model}		

We now draw connections between the approach outlined in the previous sections
with that of  the classical CMG modeling and analysis done in the Newtonian 
framework in \cite{schaub_feedback_1998}, which is
cited in much of the aerospace literature. We shall refer to this paper as the SRJ paper
henceforth. We first relate the notation and then establish a 
connection with the main equations of the SRJ paper. 

The two primary variables in the SRJ paper and ours are related as in table \ref{tab:notationcomparison}.
\begin{table}[h]
\caption{\label{tab:notationcomparison} Comparison of notation with SRJ paper}
\begin{ruledtabular}
    \begin{tabular}{ccc}
        Variable                   & This paper     & SRJ paper\\ \hline
        Gimbal angle               & $\beta$        & $\gamma$ \\
        Rotor spin magnitude       & $\dot{\gamma}$ & $\Omega$ \\
        Satellite angular velocity & $\Omega_s$     & $\omega$ \\
    \end{tabular}
\end{ruledtabular}
\end{table}

The rotation matrix in the SRJ paper, relating the gimbal and spacecraft-body frame, 
is described in terms of three orthogonal column vectors 
of unit norm, $\{ \hat{g}_s, \hat{g}_{t}, \hat{g}_g \}$, where the subscripts $s, t$ and
$g$ correspond to the {\it spin, transverse and gimbal} axes,  as
\begin{align} \pmat{| & | & |  \\  \hat{g}_s & \hat{g}_{t} & \hat{g}_g  \\ | & | & |  \\ }
\end{align}
and further,
\begin{align}
\pmat{ \inprod{\hat{g}_s}{\omega} \\  \inprod{\hat{g}_{t}}{\omega}  \\ 
    \inprod{\hat{g}_{g}}{\omega}    } = \pmat{  \omega_s  \\ \omega_t  \\  \omega_g }
\end{align} 
In our convention, the following correspondence holds: 
\begin{align} 
R_{\beta} =
\pmat{| & | & |  \\  \hat{g}_t & \hat{g}_{g} & \hat{g}_s  \\ | & | & |  \\ }
\end{align}
and
\begin{align}
R_{\beta}^T \Omega_s \longrightarrow 
\pmat{  \omega_t  \\ \omega_g  \\  \omega_s }
\end{align}
The SRJ equation of motion (eqn. 28) written partially in terms of our notation is
\begin{align}
\Itilde \dot{\Omega}_s + \hat{\Omega}_s \Itilde \Omega_s = 
\end{align}
\begin{align}
- \hat{g}_s [ J_s (\ddot{\gamma} + \dot{\beta} \omega_t)  - (J_t - J_g) \omega_t 
\dot{\beta} ]
- \hat{g}_t [ J_s( \dot{\gamma} + \omega_s) \dot{\beta} - (J_t + J_g)
\omega_s \dot{\beta}  + J_s \dot{\gamma} \omega_g ] 
- \hat{g}_g [ J_g \ddot{\beta}  -  J_s \dot{\gamma} \omega_t ]
\end{align}
while the RHS of the same equation in our notation is
\[
- [ \hat{\Omega}_s + \dot{\beta} \hat{i}_2 ]  R_{\beta} \Igimrot \dot{x}
- \dot{\beta} R_{\beta} ( \hat{i}_2 \Igimrot - \Igimrot \hat{i}_2 ) R_{\beta}^T \Omega_s
- R_{\beta} \Igimrot \ddot{x}
\]
\[
=  - \hat{g}_t [ (J_z + I_s) \dot{\gamma} \dot{\beta}  - (J_x + I_g) \dot{\beta} \omega_s
+ (J_z + I_s) \dot{\gamma} \omega_g + (J_z + I_s)  - (J_x + I_t) \dot{\beta} \omega_s]
\]
\[
- \hat{g}_g [ (J_x + I_g) \ddot{\beta} - (J_z + I_s) \dot{\gamma} \omega_t ] 		   
\] 
\[
- \hat{g}_s [ (J_z + I_s) \ddot{\gamma}  + (J_x + I_g) \dot{\beta} \omega_t  + 
((J_z + I_s) - (J_x + I_t) ) \dot{\beta} \omega_t ]
\]
\\ \vspace{1in}
The terms in the model expand as shown below.
\begin{align*}
R_{\beta} \Igimrot [0,\ddot{\beta},\ddot{\gamma}]^{T} & =\vec{g}_{g}(J_{x}+I_{g})\ddot{\beta}+\vec{g}_{s}(J_{z}+I_{s})\ddot{\gamma}\\
R_{\beta}\hat{i}_{2}\dot{\beta} \Igimrot [0,\dot{\beta},\dot{\gamma}]^{T} & =\vec{g}_{t}(J_{z}+I_{s})\dot{\gamma}\dot{\beta}\\
\hat{\Omega}_{s}R_{\beta} \Igimrot [0,\dot{\beta},\dot{\gamma}]^{T} & =\begin{array}{c}
\vec{g}_{s}((J_{x}+I_{g})\dot{\beta}\omega_{t})+\vec{g}_{g}(-(J_{z}+I_{s})\dot{\gamma}\omega_{t})\\
\vec{g}\left(-(J_{x}+I_{g})\dot{\beta}\omega_{s}+(J_{z}+I_{s})\dot{\gamma}\omega_{g}\right)
\end{array}\\
\hat{i}_{2} \Igimrot - \Igimrot \hat{i}_{2} & =
\begin{bmatrix}0 & 0 & \left(J_{z}+I_{s}-(J_{x}+I_{t})\right)\\
0 & 0 & 0\\
\left(J_{z}+I_{s}-(J_{x}+I_{t})\right) & 0 & 0
\end{bmatrix}\\
R_{\beta}(\hat{i}_{2} \Igimrot - \Igimrot \hat{i}_{2})R_{\beta}^{T}\Omega_{s}\dot{\beta} & =(\vec{g}_{t}\omega_{s}+\vec{g}_{s}\omega_{t})(\left\{ J_{z}+I_{s}-(J_{x}+I_{t})\right\} \dot{\beta})
\end{align*}

\section{Connection form}
We now detail the explicit computation of the connection form.
The principal fiber bundle structure introduces a vertical space in the tangent space at
each point on the manifold. The vertical space consists of those vectors corresponding to
the infinitesimal generator vector fields at that particular point.
The tangent space is then expressed as the direct sum of the vertical 
space and a horizontal space, where the 
horizontal space gets defined as the subspace orthogonal to the vertical space in the 
inner product induced by the Riemannian metric.
\begin{definition}

    A principal connection on $Q = (M,G,\pi)$ is a $\mathfrak{g}$ valued 1-form $\mathcal{A}$ on $Q$ satisfying, 
    \begin{enumerate}
        \item $\mathcal{A}(\xi_{Q}(q)) = \xi$, $\forall \xi \in \mathfrak{g}$ and $q \in Q$;
        \item $\mathcal{A}(\Phi_{g \ast}X_{p}) = Ad_g^* (\mathcal{A}(X_{p}))$ for all $X_{p} \in TQ$
        and for all $g \in G$. This is called the equivariance of the connection.
    \end{enumerate}
\end{definition}
The expression for the kinetic energy  is
\[
[\Omega^{T}\ u_{1}\ u_{2}]
\begin{bmatrix}
\Is + R_{\beta} \Igimrot R_{\beta}^{T} & (J_x + I_g)\vec{g} & (J_z + I_s)\vec{s}_{\beta}\\
(J_x + I_g)\vec{g}^{T} & (J_x + I_g) & 0\\
(J_z + I_s)\vec{s}_{\beta}^{T} & 0 & (J_z + I_s)
\end{bmatrix}
\begin{bmatrix}\Xi\\
w_{1}\\
w_{2}
\end{bmatrix}
\]

Action and infinitesimal generator are as shown in other sections.
So vertical space at a point 
$q$ is spanned by 
$\{(\square,0,0)|\square\in T_{q}SO(3)\}$. Locally we can represent elements in 
$T_{q}Q$ as 
$((r_{1},\ r_{2},\ r_{3}),\ \dot{\beta},\ \dot{\gamma})$ where 
$(r_{1},\ r_{2},\ r_{3})\in\mathbb{R}^{3}\simeq\mathfrak{so}(3)$.Then

\[
V_{q}Q=\mathrm{span}\{\frac{\partial}{\partial r_{1}},\ \frac{\partial}{\partial r_{2}},\ \frac{\partial}{\partial r_{3}}\}
\]

To find, 
$H_{q}Q$, the $\mathbb{G}$ orthogonal space has to be found out.

\[
\left\langle \begin{bmatrix} \Is +R_{\beta} \Igimrot R_{\beta}^{T} & (J_x + I_g)\vec{g} & (J_z + I_s)\vec{s}_{\beta}\\
(J_x + I_g)\vec{g}^{T} & (J_x + I_g) & 0\\
(J_z + I_s)\vec{s}_{\beta}^{T} & 0 & (J_z + I_s)
\end{bmatrix}\begin{bmatrix}\begin{pmatrix}h_{1}\\
h_{2}\\
h_{3}
\end{pmatrix}\\
h_{4}\\
h_{5}
\end{bmatrix},\quad\begin{bmatrix}\square\\
0\\
0
\end{bmatrix}\right\rangle =0
\]
where 
$\square\in V_{q}Q$
.
This implies
\[
(\Is +R_{\beta} \Igimrot R_{\beta}^{T})\begin{pmatrix}h_{1}\\
h_{2}\\
h_{3}
\end{pmatrix}+ (J_x + I_g) \vec{g}h_{4} + (J_z + I_s)\vec{s}_{\beta}h_{5}=0
\]

The horizontal vector 
$(h_{1}\ h_{2}\ h_{3}\ h_{4}\ h_{5})^{T}$
satisfies 3 (independent) equations in 5 variables.
The horizontal space is then 2 dimensional as expected since the number
of shape variables is 2.
If we let 
$h_{4}$ and 
$h_{5}$ be the independent variables, then
\begin{align*}
\begin{pmatrix}h_{1}\\
h_{2}\\
h_{3}
\end{pmatrix}= & (\Is +R_{\beta} \Igimrot R_{\beta}^{T})^{-1}
[(J_x + I_g)\vec{g}\ (J_z + I_s)\vec{s}_{\beta}]
\begin{pmatrix}h_{4}\\
h_{5}
\end{pmatrix}\\
= & \tilde{I}_{\beta}^{-1}
[(J_x + I_g) \vec{g}\ (J_z + I_s) \vec{s}_{\beta}]
\begin{pmatrix}h_{4}\\
h_{5}
\end{pmatrix}
\end{align*}

Now we can write any tangent vector as the sum of a vertical vector and
a horizontal vector as follows

\[
v_{q}=\begin{pmatrix}v_{1}\\
v_{2}\\
v_{3}\\
0\\
0
\end{pmatrix}+\begin{bmatrix}\tilde{I}^{-1}
[(J_x + I_g) \vec{g}\ (J_z + I_s) \vec{s}_{\beta}]\\
\begin{array}{cc}
1 & 0\\
0 & 1
\end{array}
\end{bmatrix}\begin{pmatrix}h_{4}\\
h_{5}
\end{pmatrix}
\]

The $\mathfrak{g}$ valued connection form can be then written (locally) as (here we write it as 
$\alpha:TQ\rightarrow\mathfrak{so}(3)\simeq\mathbb{R}^{3}$)

\[
\begin{bmatrix}\alpha_{11} & \alpha_{12} & \alpha_{13} & \alpha_{14} & \alpha_{15}\\
\alpha_{21} & \alpha_{22} & \alpha_{23} & \alpha_{24} & \alpha_{25}\\
\alpha_{31} & \alpha_{32} & \alpha_{33} & \alpha_{34} & \alpha_{35}
\end{bmatrix}\left(\begin{pmatrix}v_{1}\\
v_{2}\\
v_{3}\\
0\\
0
\end{pmatrix}+\begin{bmatrix}\tilde{I}_{\beta}^{-1}[(J_x + I_g) \vec{g}\ (J_z + I_s) \vec{s}_{\beta}]\\
\begin{array}{cc}
1 & 0\\
0 & 1
\end{array}
\end{bmatrix}\begin{pmatrix}h_{4}\\
h_{5}
\end{pmatrix}\right)=\begin{pmatrix}v_{1}\\
v_{2}\\
v_{3}\\
0\\
0
\end{pmatrix}
\]
solving we get,
\[
\begin{bmatrix}\alpha_{11} & \alpha_{12} & \alpha_{13} & \alpha_{14} & \alpha_{15}\\
\alpha_{21} & \alpha_{22} & \alpha_{23} & \alpha_{24} & \alpha_{25}\\
\alpha_{31} & \alpha_{32} & \alpha_{33} & \alpha_{34} & \alpha_{35}
\end{bmatrix}=\begin{pmatrix}\mathrm{Id}_{3\times3}\quad & -\tilde{I}_{\beta}^{-1}
[(J_x + I_g) \vec{g}\ (J_z + I_s)\vec{s}_{\beta}]\end{pmatrix}
\]
\section{Conclusions}
We present the spacecraft system with the variable speed control moment gyros (VSCMG)
cast into a geometric framework based on the principal fiber bundle.
The dynamics of the system are derived.
A kinematic and dynamic model for the above system is presented here.
The expressions for the associated geometric objects such as the
kinetic energy metric, locked inertia tensor,
momentum map and mechanical connection are derived.
The symmetry in the system is used to find the conserved quantity and
reduce the number of state variables in the system.
The corresponding reconstruction equations are derived.

\section*{References}
\bibliographystyle{aiaa}
\bibliography{vscmg,textbooks}
\end{document}